\documentclass[11pt, a4paper]{article}
\pdfoutput=1

\usepackage{amsmath}
\usepackage{amsfonts}
\usepackage{amssymb}
\usepackage{latexsym}
\usepackage{graphicx}
\usepackage{color}
\usepackage{mathrsfs}
\usepackage{slashed}
\usepackage{hyperref}
\usepackage{datetime}

\numberwithin{equation}{section}

\hypersetup{colorlinks, citecolor=bluscuro, linkcolor= bluscuro, urlcolor=bluscuro}
\definecolor{rossos}{cmyk}{0,1,1,0.55}
\definecolor{bluscuro}{rgb}{0.15, 0.2, .85}
\definecolor{bluchiaro}{cmyk}{1,.3,0.,0.1}
\definecolor{verdes}{rgb}{0.1, 0.5, 0.1}
\definecolor{myred}{rgb}{0.85, 0, 0}
\definecolor{myblue}{rgb}{0, 0, 0.7}
\definecolor{mygreen}{rgb}{0, 0.45, 0.1}

\makeatother   

 \def\be   {\begin{equation}}   \def\ee   {\end{equation}}
 \def\ba   {\begin{array}}      \def\ea   {\end{array}}
 \def\bea  {\begin{eqnarray}}   \def\eea  {\end{eqnarray}}
 \def\nn{\nonumber}

\def\c{\chi}

\def\g{\gamma}

\def\f{\frac}
\def\l{\left}
\def\r{\right}

\begin{document}

%
\begin{flushright} 
CERN-PH-TH/2012-369\\
SISSA 01/2013/FISI
\end{flushright}

\vspace{1.5cm}
\begin{center}

{\Huge \textbf {
On the effective operators for \\
[0.2cm]
Dark Matter annihilations
}}
\\ [1.5cm]
\textsc{
Andrea De Simone$^{1, 2, 3}$, Alexander Monin$^4$, Andrea Thamm$^{1, 4}$, Alfredo Urbano$^{2}$}
\\[1cm]

\textit{
$^1$ \textit{CERN, Theory Division, CH-1211 Geneva 23, Switzerland}\\  \vspace{1.5mm}
$^2$~SISSA, via Bonomea 265, I-34136 Trieste, Italy\\ \vspace{1.5mm}
$^3$~INFN, sezione di Trieste, I-34136 Trieste, Italy\\  \vspace{1.5mm}
$^4$ Institut de Th\'eorie des Ph\'enom\`enes Physiques,\\
 \'Ecole Polytechnique F\'ed\'erale de Lausanne, CH-1015 Lausanne,
Switzerland}
\end{center}

\vspace{1cm}
\begin{center}
\textbf{Abstract}
\begin{quote}
We consider effective 
operators describing Dark Matter (DM) interactions with Standard Model
fermions.
In the non-relativistic limit of the DM field, the operators can be organized according
to their mass dimension  and
 their velocity behaviour, i.e. whether they describe $s$-
or $p$-wave annihilations. 
The analysis is carried out for self-conjugate DM (real scalar or Majorana fermion).
In this case, the helicity suppression at work in the annihilation into fermions is lifted by
 electroweak bremsstrahlung. We construct and study all dimension-8 operators encoding
 such an effect.
These results are of interest in indirect DM searches.
\end{quote}
\end{center}

\def\thefootnote{\arabic{footnote}}
\setcounter{footnote}{0}
\pagestyle{empty}

\newpage
\pagestyle{plain}
\setcounter{page}{1}

\section{Introduction}

Significant experimental activity is currently devoted to the search for Dark Matter (DM) 
by looking at the excesses in cosmic ray production through DM annihilations (or decays) in the galactic halo.
Detailed predictions for the  fluxes largely depend on the particle physics model for the DM,
e.g. the mass, the annihilation channels etc.
It is therefore desirable to describe DM annihilations and their products within
a general and model-independent framework and Effective Field Theory (EFT) provides such a tool.

Of course, the EFT is  only applicable whenever there is a separation of scales between the process
to describe (the annihilation of non-relativistic DM at a scale $\sim M_{\rm DM}$) and the underlying
microscopic physics of the interactions (at a scale $\Lambda$). This may not always be the case, as, for example, in
the case of supersymmetry with a compressed spectrum.

If we want to describe the annihilation of two non-relativistic DM particles, whose relative velocity
is $v\sim 10^{-3}$ (in units of $c$) in our Galaxy today, it is convenient to expand the
cross section in powers of $v$ 
\be
v\sigma=a+b\,v^2+\mathcal{O}(v^4)\,,
\ee
where the first term corresponds to annihilation in the state of orbital angular momentum $L=0$ ($s$-wave)
while the second term describes $L=1$ ($p$-wave).
For the annihilation $\textrm{DM\, DM}\to f\bar f$ of a self-conjugate DM particle (real scalar or Majorana fermion)  into SM fermions of mass $m_f$,
helicity arguments lead to 
$a\propto (m_f/M_{\rm DM})^2$, and hence a very suppressed s-wave term for
light final state fermions (e.g.~leptons), while the $p$-wave is suppressed by $v^2$.

It is clear that the correct operator expansion must be done in terms of two parameters:
mass dimension of the operator and relative velocity. The effective
lagrangian would be generically given by an infinite series of non-renormalizable operators
\be
\mathscr{L}_{\rm eff}=
\sum_{d>4}
{1\over \Lambda^{d-4}}
\left(c^{(d,s)}{\cal O}^{\,(d, s)} + c^{(d,p)}{\cal O}^{\,(d, p)} \right)\,,
\ee
where ${\cal O}^{(d, s \,{\rm or}\, p )}$ indicates that the operator of dimension $d$ 
describes $s$-wave or $p$-wave annihilations,
and we neglect annihilations in waves higher than $p$.
There is no obvious ordering of the importance of the operators. An important role in this respect 
is played by ElectroWeak
(EW) bremsstrahlung, which has received significant attention recently \cite{paper0, bell1, paper1, bell2, 
cheung, ibarra1, paper2, iengo, barger, ibarra2,proc} (for earlier studies on the impact of
gauge boson radiation on DM annihilations or cosmic ray physics, see \cite{bergstrom1, bergstrom2, list1}). 
Taking into account processes with the inclusion of EW radiation  eludes the helicity suppression and 
opens up an $s$-wave contribution to the cross section 
\cite{bell1, paper1, ibarra1, paper2, ibarra2, bergstrom1}. 

In this paper we will classify the operators (up to dimension 8) according to the $v$-behaviour of the amplitude
connecting two self-conjugate particles -- real scalars or Majorana fermions --  in the initial state  with the final state of two massless fermions and possibly a gauge boson.

The effects of lifting the helicity suppression  by means of EW radiation has so far been studied within 
the context of explicit models \cite{bell1, paper1, ibarra1, paper2, ibarra2}.
The results of the present paper provide a model-independent approach to this problem and can
be used to place robust constraints on the new physics responsible for the DM sector.
Although several analyses in the literature place phenomenological constraints on the coefficients of the dimension-six operators  (see e.g. Refs.~\cite{eftconstraints1, eftconstraints2,eftconstraints3, eftconstraints4}),
the important role of higher-dimensional operators is typically underestimated.

The remainder of the paper is organized as follows. 
In Section \ref{sec:ops}, we explain our methodology and construct the 
effective operators contributing to $s$-wave DM annihilations.
We compute the differential and total annihilation cross sections
for each operator in Section \ref{sec:xsec}, and compare them
to the contribution from the typical lowest-dimensional operator.
We conclude in Section \ref{sec:conclusions}, mentioning phenomenological
applications and possible analyses that could be carried out using these results.
Finally, we collect some useful relations and identities in the Appendices, 
for the convenience of the reader.

\section{Effective operators}
\label{sec:ops}

We will carry out our analysis under the following set of assumptions on the DM sector:
\begin{enumerate}  
\item  the DM is either
a Majorana fermion or a real scalar field; 
\item the DM is neutral under the SM gauge group;
\item  there exists a $Z_2$ symmetry under which the DM is odd and the SM is even;
\item the DM couples only to the fermions in the SM spectrum, which are assumed to be massless.
\end{enumerate} 
Of course, any of these assumptions may not hold in reality and if this is the case
our analysis requires modifications. 
Assumption 1 specifies the conditions under which the helicity suppression is effective,
while  assumption 2 is there to simplify the discussion, even though it is not strictly necessary 
and relaxing this assumption can also lead to interesting effects
(see Refs.~\cite{paper2, paper3}).
Assumption 3 is commonly used in DM phenomenology to ensure the stability of the DM particle. 

More clarifications about assumption 4 are in order.
 Considering $m_f\neq 0$
would introduce another mass scale into the problem and would render the operator 
classification much less transparent. Our analysis is still valid in the regime $m_f / \Lambda \ll v$,
which may not hold for the third-generation quarks.
The other piece of information in assumption 4 is that the DM particle only couples to the fermion
sector of the SM. This needs not to be true, of course. 
Allowing for DM interactions with the other SM particles, namely gauge bosons or the Higgs,
the possibility of having additional operators contributing to  DM annihilations in 
$s$-wave opens up.
For instance for Majorana DM, $\chi$, coupling to the Higgs doublet $H$ or a generic  field-strength
$F^{\mu\nu}$ one can have $(\bar\chi\gamma^5\chi)(H^\dag H)$
and 
$\left(\bar{\chi}\gamma^5\chi\right)
F_{\mu\nu}F^{\mu\nu}$, which are CP odd, and
$\left(\bar{\chi}\gamma^5\chi\right)
F_{\mu\nu}\tilde F^{\mu\nu}$, which is CP even.
For a list of effective operators connecting DM to vector bosons and/or Higgs bosons see e.g. Refs.\cite{Rajaraman, rizzo}. We will not consider these possibilities here.

We want to classify the operators of dimension $d=6, 7,  8$ according to the $v$-behaviour of the amplitude
connecting two DM particles in the initial state  with two massless SM fermions $f$ and a gauge boson in the final state.
We look for operators which are hermitian, gauge invariant (under $SU(2)_L\otimes U(1)_Y$) and
 giving a non-zero contribution to the amplitude for the annihilation of DM into two fermions 
and one gauge boson. We will further classify the operators according to their $CP$ transformation
properties (see Appendix \ref{app:CP} for the relevant transformation properties).
The operators containing $\slashed{D}f$ give no contribution to the process under consideration,
due to the Equation Of Motion (EOM).
In order to ensure manifest gauge invariance, 
we first introduce the following notation for the covariant derivative
\begin{eqnarray}
\overrightarrow{D}_{\mu}f&=& \left(\overrightarrow{\partial}_{\mu}+igT^{a}W_{\mu}^{a}+ig^{\prime}Y_fB_{\mu}\right)f,\label{eq:cov1}\\
\bar{f}\overleftarrow{D}_{\mu}&=&
\bar{f}\left(\overleftarrow{\partial}_{\mu}-igT^{a}W_{\mu}^{a}-ig^{\prime}Y_fB_{\mu}\right),\label{eq:cov2}
\end{eqnarray} 
where $T^a=\sigma^a/2$, with $\sigma^a$ being the usual Pauli matrices, and the charge $Q_f$ of the fermion $f$
is related to its hypercharge $Y_f$ by $Q_f=T^3+Y_f$. 
Furthermore, the field-strength $W_{\mu\nu}$ of the SM gauge fields is related to the covariant derivatives as usual
\begin{eqnarray}
\left[\overrightarrow{D}_{\mu},\overrightarrow{D}_{\nu}\right]f&=&(igT^a)W^a_{\mu\nu}f\,,\label{eq:dualidentity1}\\
\overline{f}\left[\overleftarrow{D}_{\mu},\overleftarrow{D}_{\nu}\right]&=&\overline{f} (ig T^a)W^a_{\mu\nu}\,.\label{eq:dualidentity2}
\end{eqnarray}
In the following, we will separately deal with the cases where the DM is a Majorana fermion or a real scalar.
For simplicity, we will consider only left-handed SM fermions  $f_L$, but the analysis can be
applied to right-handed fermions straightforwardly.

\subsection{Majorana fermion DM}

Let us first suppose the DM particle is a Majorana fermion $\chi$.
It is possible to build several operators containing two DM fields, two SM fermion fields and zero or 
one gauge bosons.
They can be built in full generality requiring gauge invariance and hermiticity,
and further classified according to their $CP$ properties.
The Majorana-flip properties and the chiralities of the SM fermions make several structures identically zero.
The only two Majorana fermion bilinears that are non-vanishing in the limit $v\to 0$ are $\bar{\chi}\gamma^5\chi$
and $\bar\chi\gamma^\mu\gamma^5\chi$ (see  Appendix \ref{app:bilinears}).
As we are interested in $s$-waves only, we will limit our analysis to these two bilinears for the Majorana fermions.
For the pseudo-scalar bilinear  $\bar \chi \gamma^5\chi$ no contractions with SM fermions can be built,
as they would vanish either by chirality or by the 
EOM of  fermions, so we are left with the axial-vector bilinear. 

The lowest-dimensional terms that can be written out of two $\chi$'s and two $f_L$'s are of dimension 6.
At this level, there is only one non-vanishing operator satisfying all criteria:
\begin{equation}
\mathcal{O}_{\rm M}=
\left(\bar{\chi}\gamma^5\gamma^{\mu}\chi\right)
\left[
\bar{f}_L\gamma_{\mu} f_L
\right]\,.
\label{O6}
\end{equation}
The $\mu=0$ component of this operator could {\it a priori} give a $v$-independent
contribution to the scattering amplitude, but it actually vanishes because of the identity
$u_f^\dag({\mathbf p}) v_f(-{\mathbf p})=0$.
Therefore this dimension-6 operator only contributes to the $p$-wave, as  noted in Ref.~\cite{paper1}, due to the helicity suppression, which cannot be removed
by simply radiating a gauge boson from the external final state leg.
In order to look for $s$-wave terms, we need to consider higher-dimensional operators with one EW gauge
boson whose radiation in the annihilation process  lifts the helicity suppression.

At dimenion 7, there are no operators contributing to the $s$-wave cross section.
In fact, all possible structures vanish either due to Majorana-flip properties or the chirality of the SM fermions, or because of the EOM of the $f_L$s.
A priori, the $\mu=0$ component of the operator $(\bar{\chi}\gamma^5{\partial_\mu}\chi)
[\overline{f}_L\gamma^\mu f_L]$ would give a $v$-independent contribution to the 
scattering amplitude, but it vanishes again because of the identity
$u_f^\dag({\mathbf p}) v_f(-{\mathbf p})=0$. 

At the level of dimension 8, there are several structures that can be built 
requiring gauge invariance and hermiticity. 
They contain two $\chi$'s (in the bilinear $\bar\chi\gamma^\mu\gamma^5\chi$), two $f_L$'s and two covariant derivatives. 
It is possible to reduce the number of independent operators, contributing to the
cross section for the process under consideration, by using EOM and the identities 
in Appendix \ref{app:identities}.
In addition, some structures can be related to each other by  terms (like 
$\left(\bar{\chi}\gamma^5\gamma^{\mu}\chi\right)\partial^2\left[
\bar{f}_L\gamma_{\mu}f\right]$),
which do not contribute to the $s$-wave annihilation into two fermions and a gauge boson.
Therefore they contribute in exactly the same way to the amplitude for the process we are interested in. 
There remain only five independent operators of dimension 8 contributing to the 
 $s$-wave annihilation of DM into two SM fermions and a gauge boson,
 listed in Table \ref{tableOM} together with their $CP$ conjugation properties.
We remain agnostic about the presence or absence of $CP$ violation in the Dark Matter
sector, which can possibly induce $CP$ violation in the SM at loop level and therefore be
further constrained.
\begin{table}
\begin{center}
\begin{tabular}{|c|c|c|}
\hline
\textbf{Name} & \textbf{Operator} & \textbf{CP}\\
\hline
$\mathcal{O}_{{\rm M}1}$ & 
$\left(\overline{\chi}\gamma^5\gamma^{\mu}\chi\right)
\left[\left(\overline{f}_L\overleftarrow{D}_{\rho}\right)\gamma_{\mu}\left(\overrightarrow{D}^{\rho}f_L\right)\right]$ & $+$\\
\hline
$\mathcal{O}_{{\rm M}2}$& 
$i \epsilon_{\mu\nu\rho\sigma} \left(\overline{\chi}\gamma^5\gamma^{\mu}\chi\right)
\left[\overline{f}_L\gamma^{\nu}\overrightarrow{D}^{\rho}\left(\overrightarrow{D}^{\sigma}f_L\right)-
\left(\overline{f}_L
\overleftarrow{D}^{\sigma}\right)\overleftarrow{D}^{\rho}\gamma^{\nu}f_L\right]$
& $+$ \\
\hline
$\mathcal{O}_{{\rm M}3}$ & 
$i\epsilon_{\mu\nu\rho\sigma}
\left(\overline{\chi}\gamma^5\gamma^{\mu}\chi\right)
\left[\left(\overline{f}_L\overleftarrow{D}^{\nu}\right)\gamma^{\rho}
\left(\overrightarrow{D}^{\sigma}f_L\right)-
\left(\overline{f}_L
\overleftarrow{D}^{\sigma}\right)\gamma^{\rho}\left(\overrightarrow{D}^{\nu}f_L\right)\right]$
& $+$\\
\hline
$\mathcal{O}_{{\rm M}4}$ &
$
i \left(\overline{\chi}\gamma^5\gamma^{\mu}\chi\right) \left[\overline{f}_L\overrightarrow{\slashed{D}}\left(\overrightarrow{D}_{\mu}f_L\right)-\left(\overline{f}_L
\overleftarrow{D}_{\mu}\right)\overleftarrow{\slashed{D}}f_L\right]$
& $-$ \\
\hline
$\mathcal{O}_{{\rm M}5}$ & 
$ i \left(\overline{\chi}\gamma^5\gamma^{\mu}\chi\right) \left[\overline{f}_L\gamma_{\mu}\overrightarrow{D}_{\rho}\left(\overrightarrow{D}^{\rho}f_L\right)-\left(\overline{f}_L
\overleftarrow{D}_{\rho}\right)\overleftarrow{D}^{\rho}\gamma_{\mu}f_L\right]$
& $-$\\
\hline
\end{tabular}
\caption{List of dimension-8 operators contributing to the $s$-wave  cross section for the
annihilation of Majorana DM into two fermions and a gauge boson.
\label{tableOM}}
\end{center}
\end{table}

 All other operators have either a larger dimensionality or produce more powers of $v^2$ in
the annihilation cross section.
Notice that we chose to keep a Lorentz-covariant formalism, despite looking at the non-relativistic
limit. This implies that the same operator can lead to both $v$-independent and $v$-dependent
terms in the amplitudes; for example, the operator $\mathcal{O}_{{\rm M}1}$ also gives a contribution
to the $p$-wave cross section, but we will not consider it as it is very suppressed.

To summarize, the dimension-8 operator contributing to the $s$-wave annihilation
cross section of Majorana DM into SM fermions is given by the sum of the operators in Table
\ref{tableOM},
$\mathcal{O}_{{\rm M}}^{(8,s)}=\sum_{i=1}^5 c^{(8,s)}_{i} \mathcal{O}_{{\rm M}i}$;
the leading interactions are therefore described in terms of only a few  operators
\be
\mathscr{L}_{\rm eff}=
{1\over \Lambda^2}c^{(6,p)}{\cal O}_{\rm M}+
{1\over \Lambda^4}\sum_{i=1}^5 c^{(8,s)}_{i}  \mathcal{O}_{{\rm M}i}
+\textrm{higher-dim}\,.
\ee
In absence of $CP$-violation in the DM sector, only the first three operators
in the sum need to be considered.

\subsection{Real scalar DM}
\label{subsec:scalarops}

Next, let us consider the case where the DM particle is a real scalar $\phi$.
By angular momentum conservation, two real scalars  cannot annihilate into two massless fermions in the configuration with $L=0$; this process would  only occur through a chirality flip induced by a mass term. For this reason,
in the limit $m_f=0$ we are considering here, no $s$-wave annihilation $\phi\phi \to \bar f_L f_L$ is possible.
One can recover this result in the language of effective operators.
At dimension 5, only a single operator can be constructed which, however, vanishes due to chirality, 
$\phi^2 (\overline{f}_L f_L) = 0$.
At dimension 6, we have $\phi^2 \partial_\mu(\overline{f}_L\gamma^\mu f_L) = 0$,
 by the EOM (cf. Eq.~(\ref{eq:ID1bis})).
At dimension 7, there are no possible Lorentz contractions to construct an operator.

Nevertheless,
at the level of dimension 8, several gauge invariant hermitian operators can be built out of two
$\phi$'s, 
two $f_L$'s and covariant derivatives. As discussed already for the Majorana case, 
it is possible to reduce the number of independent operators, contributing to the
cross section for the process under consideration, by using the EOM and the identities 
in Appendix \ref{app:identities}.  
We are left with four CP-even operators and three CP-odd operators,  listed in Table \ref{tableOR}.
A $v$-dependent  annihilation $\phi\phi \to \bar f_L f_L$ is mediated by the operator ${\cal O}_{\rm R4}$, while the $s$-wave annihilation of two $\phi$'s can proceed by switching on the emission
of a gauge boson in the final state.

Other operators involving $\phi\partial_\mu\phi$ or a gauge field-strength $F^{\mu\nu}$ can be obtained from the listed ones by integration by parts, using the EOM of the DM particle or the identities
(\ref{eq:dualidentity1})-(\ref{eq:dualidentity2}).
For instance, the operator $\phi^2 \partial_\nu \left[ \bar{f}_L \gamma_\mu f_L \right] F^{\mu\nu}$, considered in Ref.~\cite{barger2}, is expressed in this basis as $1/g ({\cal O}_{\rm R1}-{\cal O}_{\rm R2})$.

\begin{table}
\begin{center}
\begin{tabular}{|c|c|c|}
\hline
\textbf{Name} & \textbf{Operator} & \textbf{CP}\\
\hline
$\mathcal{O}_{{\rm R1}}$ & 
$i \phi^2 \left[\overline{f}_L\overrightarrow{\slashed{D}}\overrightarrow{D}^{\nu}
 (\overrightarrow{D}_{\nu}f_L)-(\overline{f}_L\overleftarrow{D}_{\nu})\overleftarrow{D}^{\nu}\overleftarrow{\slashed{D}}f_L
 \right]
$ & $+$\\
\hline
$\mathcal{O}_{{\rm R2}}$ & 
$i\phi^2 \left[\overline{f}_L\overrightarrow{D}^{\nu}\overrightarrow{\slashed{D}}
 (\overrightarrow{D}_{\nu}f_L)-(\overline{f}_L\overleftarrow{D}_{\nu})\overleftarrow{\slashed{D}}\overleftarrow{D}^{\nu}f_L
 \right]
$ & $+$\\
\hline
$\mathcal{O}_{{\rm R3}}$ & 
$ \phi^2 \epsilon_{\mu\nu\rho\sigma} \left[
   \overline{f}_L\gamma^{\mu}\overrightarrow{D}^{\nu}\overrightarrow{D}^{\rho}(\overrightarrow{D}^{\sigma}f_L)+
   (\overline{f}_L\overleftarrow{D}^{\sigma})\overleftarrow{D}^{\rho}\overleftarrow{D}^{\nu}\gamma^{\mu}f_L
   \right] 
$ & $+$\\
\hline
$\mathcal{O}_{{\rm R4}}$ &
$i (\partial_\mu\phi\partial_\nu\phi)
\left[\bar f_L\gamma^\mu
\overrightarrow{D}^{\nu} f_L-
\bar f_L\overleftarrow{D}^{\nu}\gamma^\mu f_L\right]$
& $+$\\
\hline
$\mathcal{O}_{{\rm R5}}$& 
$\phi^2 \left[(\overline{f}_L\overleftarrow{D}^{\nu})\overrightarrow{\slashed{D}}
 (\overrightarrow{D}_{\nu}f_L)+(\overline{f}_L\overleftarrow{D}_{\nu})\overleftarrow{\slashed{D}}(\overrightarrow{D}^{\nu}f_L)
 \right]$
& $-$ \\
\hline
$\mathcal{O}_{{\rm R6}}$ & 
$i \phi^2 \epsilon_{\mu\nu\rho\sigma} \left[
   \overline{f}_L\gamma^{\mu}\overrightarrow{D}^{\nu}\overrightarrow{D}^{\rho}(\overrightarrow{D}^{\sigma}f_L)-
   (\overline{f}_L\overleftarrow{D}^{\sigma})\overleftarrow{D}^{\rho}\overleftarrow{D}^{\nu}\gamma^{\mu}f_L
   \right]$
& $-$\\
\hline
$\mathcal{O}_{{\rm R7}}$ &
$
i \phi^2\epsilon_{\mu\nu\rho\sigma}\left[
   (\overline{f}_L\overleftarrow{D}^{\nu})\gamma^{\mu}\overrightarrow{D}^{\rho}(\overrightarrow{D}^{\sigma}f_L)-
   (\overline{f}_L\overleftarrow{D}^{\sigma})\overleftarrow{D}^{\rho}\gamma^{\mu}(\overrightarrow{D}^{\nu}f_L)
   \right]$
& $-$ \\
\hline
\end{tabular}
\caption{List of dimension-8 operators contributing to the $s$-wave  cross section for the
annihilation of real scalar DM into two fermions and a gauge boson.
\label{tableOR}}
\end{center}
\end{table}

To summarize, the dimension-8 operator contributing to the $s$-wave annihilation
cross section of real scalar DM into SM fermions is given by the sum of the operators in Table 
\ref{tableOR}:
$\mathcal{O}_{{\rm R}}^{(8,s)}=\sum_{i=1}^7 c^{(8,s)}_{i}  \mathcal{O}_{{\rm R}i}$;
the leading interactions are therefore described in terms of only a few  operators
\be
\mathscr{L}_{\rm eff}=
{1\over \Lambda^4}\sum_{i=1}^7 c^{(8,s)}_{i} \mathcal{O}_{{\rm R}i}
+\textrm{higher-dim}\,.
\ee
In absence of $CP$-violation in the DM sector, only the first four operators
in the sum need to be considered.

\section{Annihilation cross sections}
\label{sec:xsec}

\begin{figure}[t]
\begin{center}
  \includegraphics[width=12 cm]{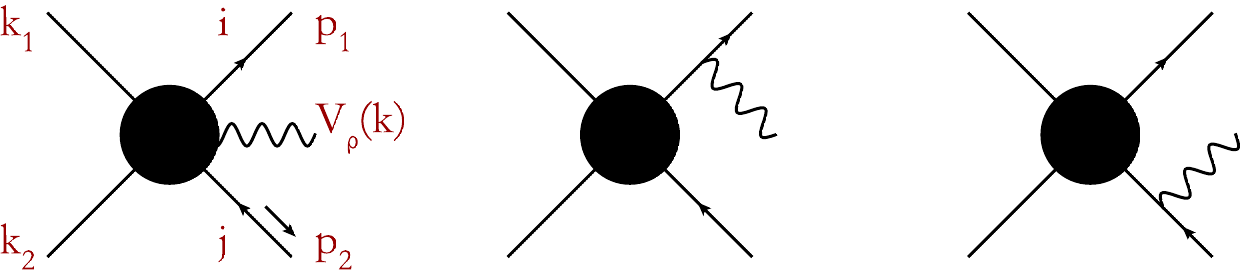}
  \caption{\emph{Diagrams for the annihilation process in Eq.~(\ref{eq:MainAnnihilation}).}}\label{fig:EffectiveAmplitude}
  \end{center}
\end{figure}

In this section we show analytical results for the annihilation cross sections due to the operators
found above.
For simplicity, we restrict ourselves to considering left-handed SM fermions only, but the results can be 
easily adapted to account for annihilations into right-handed fermions as well.
We consider  the process
\begin{equation}\label{eq:MainAnnihilation}
{\rm DM}(k_1)\, {\rm DM}(k_2)\to 
f_{i,L}(p_{1})\, \bar{f}_{j,L}(p_2)\, V(k),
\end{equation}
where light fermions in the final state - described here by the generic $SU(2)_L$ doublet $F=(f_1,f_2)^{T}$  - can be both leptons and quarks, and where $V=W^{\pm},Z,\gamma$. 
Note that diagrams with gauge boson emission from the final state legs have to be included in order to compute a gauge invariant amplitude (see Fig.~\ref{fig:EffectiveAmplitude}).

It is convenient to  introduce the kinematical variables $y, z$ defined by
\begin{eqnarray}
 p_1^0 &=& (1-y)\sqrt{s}/2\, ,\label{eq:p1} \\
 p_2^0 &=& (1-z)\sqrt{s}/2\, ,\label{eq:p2}\\
  k^0 &=& (y+z)\sqrt{s}/2\, , \label{eq:k}
\end{eqnarray}
which are subject to the following phase space constraints
\be
\frac{m_{V}^{2}}{s} \leq  y \leq 1\,,\quad
\frac{m_{V}^{2}}{s \, y} \leq  z \leq 1 - y + \frac{m_{V}^{2}}{s} \,,
\label{eq:phasespace}
\ee
where $s=(k_1+k_2)^2= 4 M_{\rm DM}^2 / (1 - v^2/4)$.
The scattering amplitudes will be proportional to a coefficient $A$ containing
the correct gauge couplings according to the different possible final states and their $SU(2)_L\otimes U(1)_Y$ quantum numbers; more explicitly,
\be
\begin{tabular}{rclrcl}
$A(f_{1}f_{1}\gamma)$ &=& $- \dfrac{g s_W}{2}(1+y_f)$\,,
&
$A(f_2f_2\gamma) $ &=& $+\dfrac{g s_W}{2}(1-y_f)$\,,\\
$A(f_1f_1Z)$ &=& $-\dfrac{g}{2c_W}[1-(1+y_f)s_W^2]$\,,
&
$A(f_2f_2Z)$ &=&  $+\dfrac{g}{2c_W}[1-(1-y_f)s_W^2]$\,,\\
$A(f_1f_2W^+)$ &=& $-\dfrac{g}{\sqrt{2}}$\,,
&
$A(f_2f_1W^-)$ &=& $-\dfrac{g}{\sqrt{2}}$\,,
\end{tabular}
\ee
where $Y_f=y_f/2$ (e.g. $y_L=-1$, $y_Q=1/3$), $g$ is the gauge coupling 
and $s_W, c_W$ are the sin and cos of the
weak angle, respectively.

The double-differential annihilation cross section is given in terms of the 
squared amplitude $\overline{|\mathcal{M}|^2}$ (averaged
over the initial spins and summer over the final ones) as
\begin{equation}
v\frac{d^2\sigma}{dy dz}={\overline{|\mathcal{M}|^2}\over 128 \, \pi^3}\,,
\end{equation}
while the total cross section is obtained by integrating over the kinematical
variables on the phase space domain defined by Eq.~(\ref{eq:phasespace}).
Let us now show the results of the computation of these cross sections 
 for each of the operators found in the previous section.

\subsection{Majorana fermion DM}

In the zero-velocity approximation, $s=4M_{\rm DM}^2$ and the double-differential cross sections
for the annihilation  process (\ref{eq:MainAnnihilation}) mediated by the dimension-8  operators 
$\mathcal{O}_{{\rm M}1}, \ldots, \mathcal{O}_{{\rm M}5}$ 
in Table \ref{tableOM} are
\begin{eqnarray}
\left.v\frac{d^2\sigma}{dydz}\right\vert_{{\cal O}_{{\rm M}1}} &=& |c_{{\rm M}1}^{(8, s)}|^{2}
\frac{A^2\, M_{\rm DM}^6} {2 \pi^{3} \Lambda^{8}}
\left[
1-y-z+\frac{m_V^2}{4M_{\rm DM}^2}
\right]\left[
y^2+z^2-\frac{m_V^2}{2M_{\rm DM}^2}
\right]\,,\\
\left.v\frac{d^2\sigma}{dydz}\right\vert_{{\cal O}_{{\rm M}2}}  &=& 
4 \, \frac{|c_{{\rm M2}}^{(8, s)}|^{2}}{|c_{{\rm M1}}^{(8, s)}|^{2}} 
\left.v\frac{d^2\sigma}{dydz}\right\vert_{{\cal O}_{{\rm M}1}} 
 \,,\\
\left.v\frac{d^2\sigma}{dydz}\right\vert_{{\cal O}_{{\rm M}3}}  &=& 
4 \, \frac{|c_{{\rm M}3}^{(8, s)}|^{2}}{|c_{{\rm M}1}^{(8, s)}|^{2}} 
\left.v\frac{d^2\sigma}{dydz}\right\vert_{{\cal O}_{{\rm M}1}} 
\,,\\
\left.v\frac{d^2\sigma}{dydz}\right\vert_{{\cal O}_{{\rm M}4}}  &=& |c_{{\rm M}4}^{(8, s)}|^{2}
\frac{2 A^2\,M_{\rm DM}^6}{\pi^{3} \Lambda^{8}} \left[
(1-y-z)(y^2+z^2)+\frac{m_V^2}{4M_{\rm DM}^2}\left[(y+z)^2-2(y+z)+2\right]
\right]
\,,\nn\\
&&\\
\left.v\frac{d^2\sigma}{dydz}\right\vert_{{\cal O}_{{\rm M}5}}  &=&
\frac{|c_{{\rm M}5}^{(8, s)}|^{2}}{|c_{{\rm M}4}^{(8, s)}|^{2}}
\left.v\frac{d^2\sigma}{dydz}\right\vert_{{\cal O}_{{\rm M}4}}  
\,.
\end{eqnarray}
Notice that in the limit $m_V/M_{\rm DM}\to 0$ the differential cross sections
are the same for all operators, up to an overall numerical factor.

\begin{figure}[t]
\begin{center}
\includegraphics[width=7.5cm]{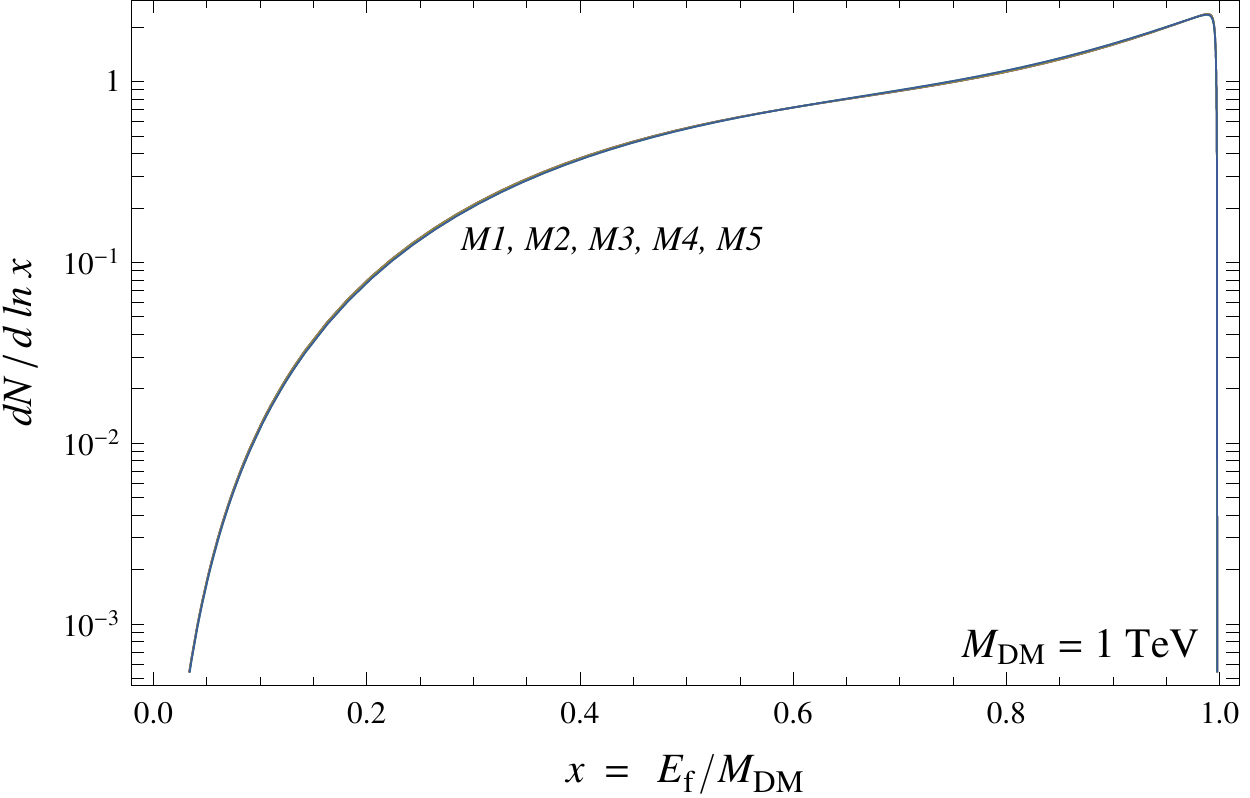}
\hspace{0.5cm}
\includegraphics[width=7.5cm]{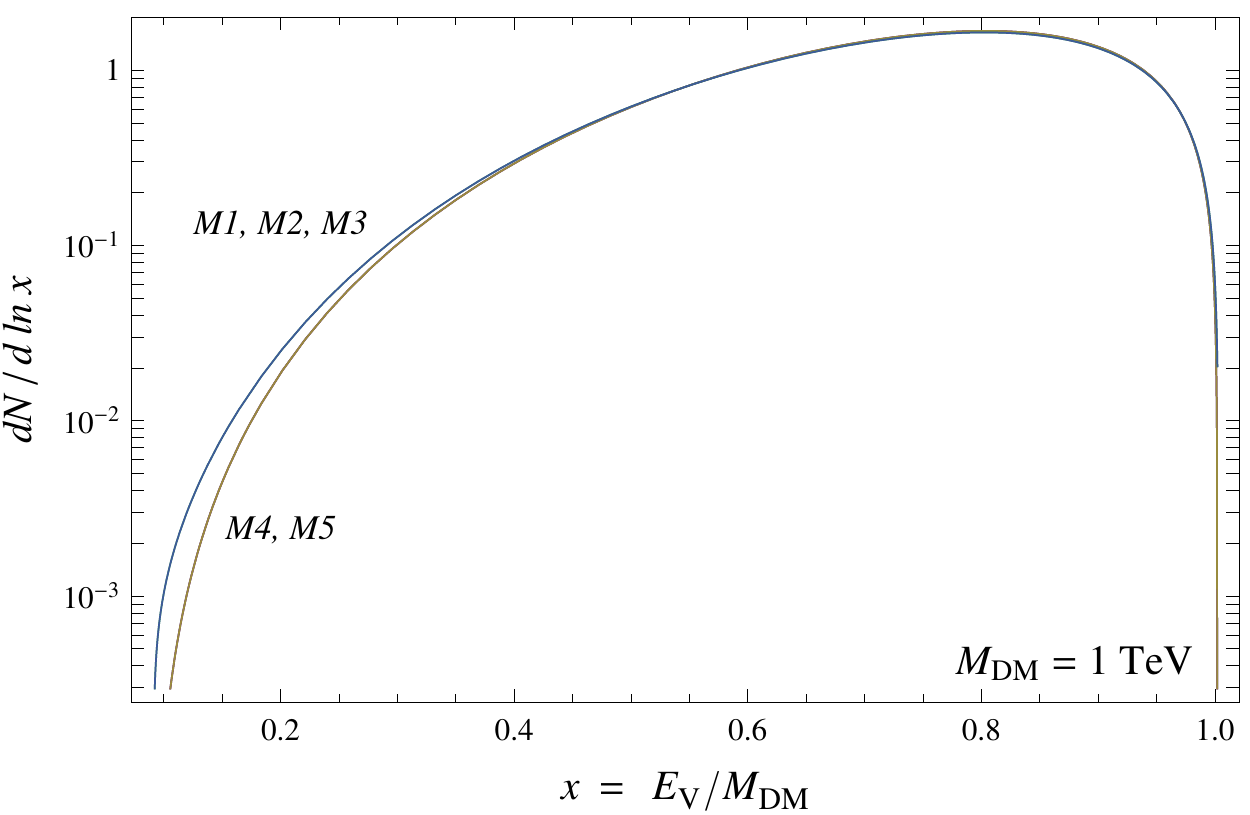}
  \caption{Majorana fermion DM. \emph{The energy distributions $d N/d\ln(E/M_{\rm DM})$ of the final fermion \emph{(left panel)} and of the final gauge boson \emph{(right panel)},  for the different dimenion-8 operators.
We  set $c^{(6,p)}=c^{(8,s)}_{{\rm M}i}=1$ and $M_{\rm DM}=1$ TeV.
 }}
\label{fig:spectraM}
  \end{center}
\end{figure}

In the indirect searches for DM, the observables measured experimentally 
are the fluxes of cosmic rays, which are directly related to the energy spectra of particles
generated by DM annihilations at the production point. By integrating the double-differential cross sections listed above once, one obtains the energy spectra of the SM fermions
and of the gauge bosons at production. 
We consider the  distributions of the final fermion energy ($E_f$) and of the final gauge boson energy ($E_V$), defined as 
\be
\frac{dN}{d\ln x}\equiv \frac{1}{\sigma(\textrm{DM\,DM} \to f \overline f V)}\frac{d\sigma(\textrm{DM\,DM} \to f \overline f V) }{d\ln x}\,,
\label{dNdx}
\ee
where $x\equiv E_{f, V}/M_{\rm DM}$, and shown in Fig.~\ref{fig:spectraM}.

The distributions originating from the set of operators ${\cal O}_{\rm M1}, {\cal O}_{\rm M2}, {\cal O}_{\rm M3}$ differ just by an overall factor, which cancels out by normalizing the spectra, and therefore produce the same curves.
The same argument applies to the other set ${\cal O}_{\rm M4}, {\cal O}_{\rm M5}$.
However, the operator ${\cal O}_{\rm M1}$ would instead give a different result with respect to ${\cal O}_{\rm M4}$, but the difference is not visible in the fermion energy distribution (left panel of Fig.~\ref{fig:spectraM})
because of the smallness of $m_V/M_{\rm DM}$.

The dimension-6 operator ${\cal O}_{\rm M}$ mediates the  two-body annihilation
$\chi\chi \to f_L\overline{f}_L$ in $p$-wave.
The processes where a gauge boson is radiated from the final state fermion
are still in $p$-wave, 
as discussed in Ref.~\cite{paper1}. Therefore, the energy spectra of fermions and gauge bosons originating from the dimension-6 operator are going to be much less important than those from the dimension-8
operators.
The total cross section for  the two-body process $\chi\chi \to f_L\overline{f}_L$ mediated by ${\cal O}_{\rm M}$ is simply
\be
\left.v\sigma(\chi\chi \to f_L\overline{f}_L)\right\vert_{{\rm M}}=  |c^{(6,p)}|^2 \frac{  M_{\rm DM}^{2}}{12 \pi \Lambda^{4}} 
v^{2}\,,
\ee
while the total cross sections of the three-body processes (\ref{eq:MainAnnihilation}) mediated by
the dimension-8 operators are obtained by integrating over the full phase space 
\bea
\left.v\sigma\right\vert_{{\rm M}1}&=&|c_{{\rm M}1}^{(8, s)}|^{2} \frac{ A^{2} M_{\rm DM}^{6}}{240 \pi^{3} \Lambda^{8}} \left[ 4  - 15 \frac{m_{V}^{2}}{M_{\rm DM}^{2}} + 5 \frac{m_{V}^{4}}{M_{\rm DM}^{4}} \left( -4 + 6 \ln \frac{2 M_{\rm DM}}{m_{V}}  \right)
+\mathcal{O}\left(\frac{m_V^6}{M_{\rm DM}^6}\right) \right]\,, \\
\left.v\sigma\right\vert_{{\rm M}2}&=& 
4 \,\frac{|c_{{\rm M}2}^{(8, s)}|^{2}}{|c_{{\rm M}1}^{(8, s)}|^{2}} 
\left.v\sigma\right\vert_{{\rm M}1}\,, \\
\left.v\sigma\right\vert_{{\rm M}3}&=&
4 \, \frac{|c_{{\rm M}3}^{(8, s)}|^{2}}{|c_{{\rm M1}}^{(8, s)}|^{2}} 
\left.v\sigma\right\vert_{{\rm M}1} \,, \\
\left.v\sigma\right\vert_{{\rm M}4}&=& |c_{{\rm M}4}^{(8, s)}|^{2} \frac{ A^{2} M_{\rm DM}^{6}}{120 \pi^{3} \Lambda^{8}} \left[8 + 15 \frac{m_{V}^{2}}{M_{\rm DM}^{2}} + \frac{m_{V}^{4}}{M_{\rm DM}^{4}}  \left( 40 - 60 \ln \frac{2 M_{\rm DM}}{m_{V}} \right)+\mathcal{O}\left(\frac{m_V^6}{M_{\rm DM}^6}\right) \right] \,,\\
\left.v\sigma\right\vert_{{\rm M}5}&=&
\frac{|c_{{\rm M}5}^{(8, s)}|^{2}}{|c_{{\rm M}4}^{(8, s)}|^{2}}
\left.v\sigma\right\vert_{{\rm M}4}\,.
\eea
For simplicity, we have only reported here the leading terms in the expansion in powers of 
$m_V/M_{\rm DM}$,
but the complete analytical expressions are used in the plots.
The sub-leading terms can  be of the same order as the contributions from higher-dimensional
operators we are neglecting.

\begin{figure}[t]
\begin{center}
 \includegraphics[width=7.5cm]{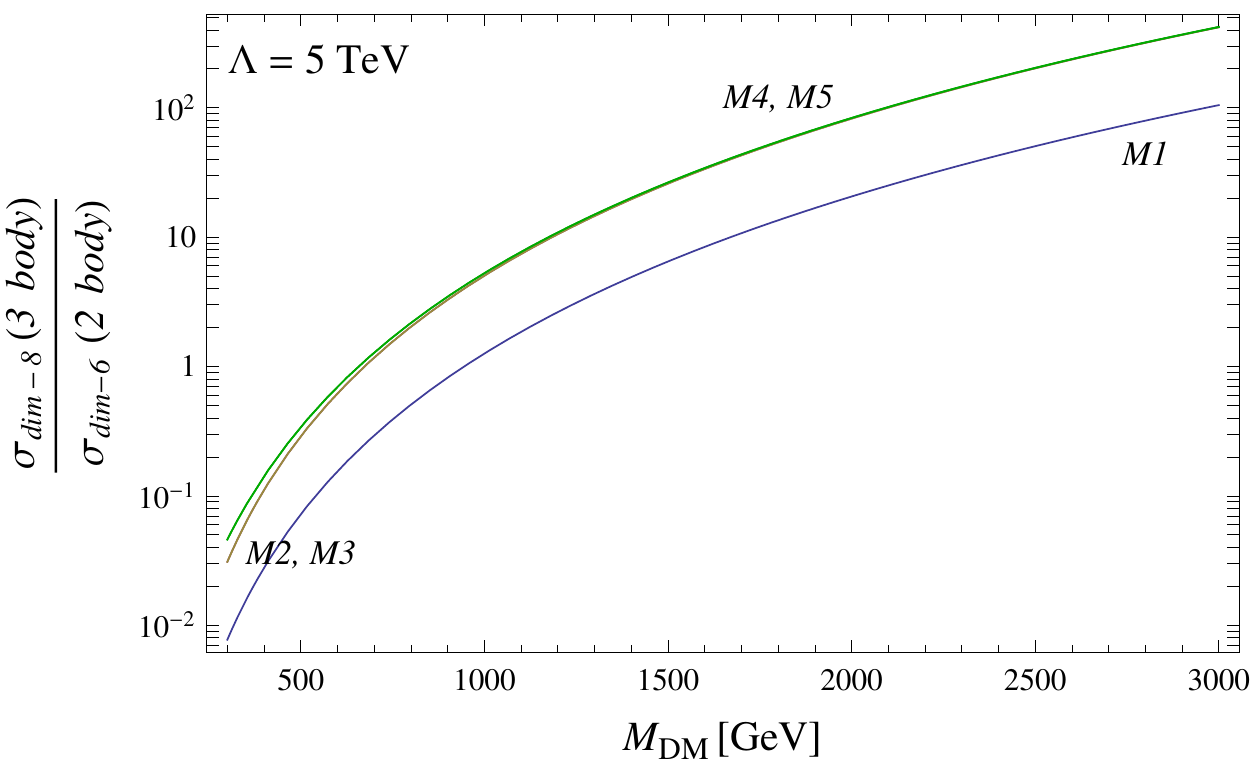}
 \hspace{0.5cm}
 \includegraphics[width=7.5cm]{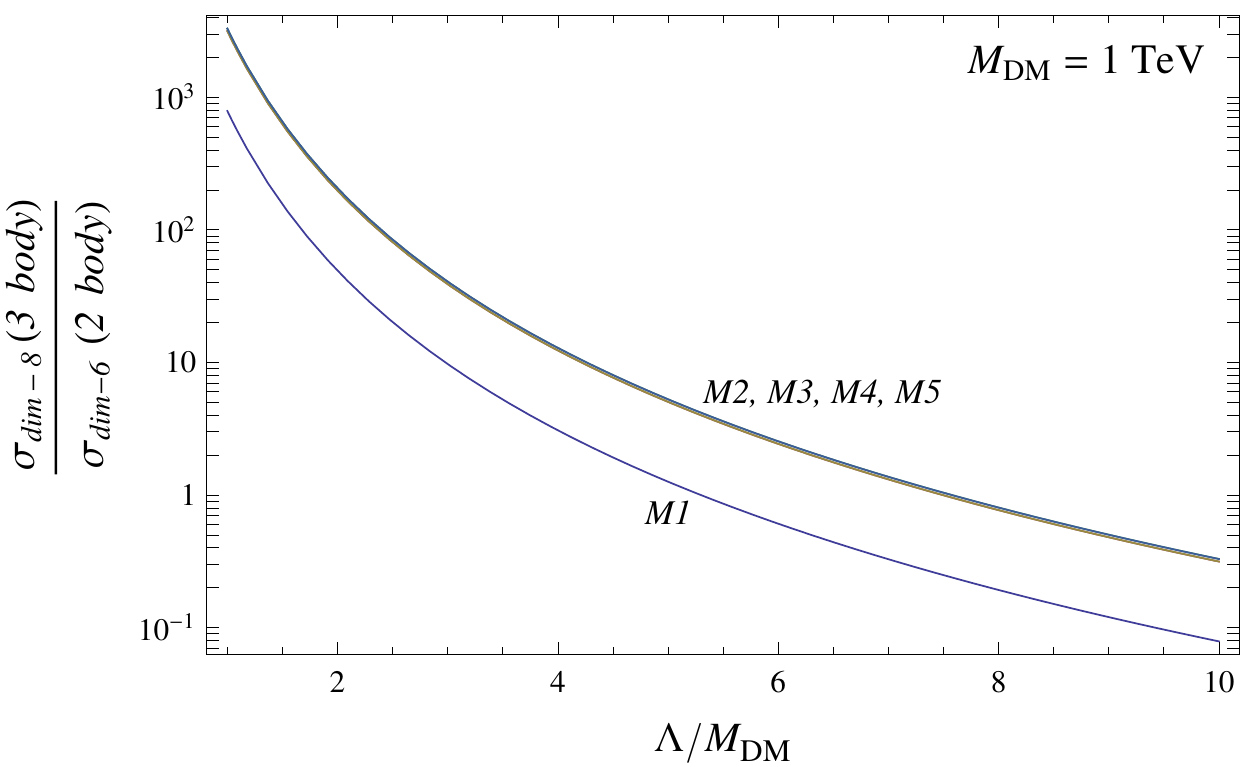}
  \caption{Left panel: \emph{The ratio 
  of the total cross sections $\left.v\sigma(\chi\chi\to \ell_L^+\ell_L^-Z)\right\vert_{{\rm Mi}}/
  \left.v\sigma(\chi\chi \to \ell_L^+\ell_L^-)\right\vert_{{\rm M}}$
as a function of $M_{\rm DM}$. We set $c^{(6,p)}=c^{(8,s)}_i=1$, and $\Lambda=5$ TeV. 
The curves corresponding to ${\cal O}_{\rm M4}, {\cal O}_{\rm M5}$ (in green) departs from those
of ${\cal O}_{\rm M2}, {\cal O}_{\rm M3}$ at low $M_{\rm DM}$ and are slightly higher.}
Right panel: \emph{The ratio 
  of the total cross sections $\left.v\sigma(\chi\chi\to \ell_L^+\ell_L^-Z)\right\vert_{{\rm Mi}}/
  \left.v\sigma(\chi\chi \to \ell_L^+\ell_L^-)\right\vert_{{\rm M}}$
as a function of  $\Lambda/M_{\rm DM}$.
We  set $c^{(6,p)}=c^{(8,s)}_i=1$ and $M_{\rm DM}=1$ TeV. }}
\label{fig:ratioM}
  \end{center}
\end{figure}

The relative importance of the $s$-wave three-body process due to dimension-8 operators with respect
to the $p$-wave two-body annihilation due to the dimenion-6 operator is captured by the
ratios of the total cross sections,  plotted in Fig.~\ref{fig:ratioM}.
Three-body cross sections can be sizeably larger than the two-body ones.
 It is evident that the dimension-8 operators dominate the total cross section,
provided  that the effective operator scale $\Lambda$ is not too large with respect to the DM mass $M_{\rm DM}$.
It is clear that limiting an EFT analysis for DM annihilations to the dimension-six operators
 misses the right result, as the cross section receives important contributions from operators
of dimension higher than six.

One could have expected this result by an order-of-magnitude estimate
 of the two-body and three-body total cross sections originating from the 
operators above
\bea
\left.v\sigma(2\to 2)\right\vert_{{\cal O}_6} &\sim&  |c^{(6,p)}|^2 {v^2}{ M_{\rm DM}^2\over \Lambda^4}\\
\left.v\sigma(2\to 3)\right\vert_{{\cal O}_8} &\sim& |c^{(8,s)}|^2 {\alpha_W\over 4\pi}  
{ M_{\rm DM}^6\over \Lambda^8}\,,
\eea
where $\alpha_W=g^2/(4\pi)$,
from which we learn that the three-body cross section due to the dimension-8 operator can be bigger than
 the two-body one due to dimension-6, provided that
\be
{|c^{(6,p)}|\over |c^{(8,s)}|}
{\Lambda^2\over M_{\rm DM}^2}\lesssim {1\over v} \sqrt{{\alpha_W\over 4\pi}}\simeq 50
\label{estimate}
\ee
for $v=10^{-3}$. 
This estimate is confirmed by the numerical results in Fig.~\ref{fig:ratioM}.

Let us conclude this subsection with a comment on  the toy-model studied in Ref.~\cite{paper1}, consisting of a Majorana spinor $\chi$ with mass $M_{\rm DM}$ and a scalar $S$ with mass $M_{S} > M_{\rm DM}$, in addition to the SM particle content. The added particles are a SM singlet and a $SU(2)_L$-doublet respectively. The Lagrangian is of the form
\begin{equation}
\mathcal{L} = \mathcal{L}_{\rm SM} +  \frac{1}{2} \bar{\chi} ( i \slashed{\partial} - M_{\rm DM} ) \chi + 
(D_{\mu} S)^{\dagger}(D_{\mu}S) - M_S^{2} S^{\dagger} S +(y_{\chi} \bar{\chi} (F i \sigma_{2} S) + \textrm{h.c.})\,.
\label{toymodel}
\end{equation}
In the limit $M_S\gg M_{\rm DM}$, the interactions can be accurately described by the effective operator
${\cal O}_{{\rm M}1}$ in Table \ref{tableOM}.
Indeed, in  the limit $v \rightarrow 0$, the amplitudes for the process $\chi\chi\to f_L \bar{f}_L Z$
as due to the toy model  Eq.~(\ref{toymodel}) and the effective operator ${\cal O}_{{\rm M}1}$ match, provided that $4 c^{(8,s)}_{ \rm{M} 1}/\Lambda^4= -  y_\chi^2/M_S^4$.

\subsection{Real scalar DM}

As discussed in the previous section, there is no two-body annihilation $\phi\phi \to  f_L\overline{f}_L$
in $s$-wave
in the limit $m_f=0$. So the EW bremsstrahlung opens up a three-body annihilation channel
which is otherwise absent. The first non-vanishing contribution to the $s$-wave annihilation cross section
comes from the dimension-8 operators  in Table \ref{tableOR} and 
there is no contribution from lower-dimensional operators to compare with.

\begin{figure}[t]
\begin{center}
\includegraphics[width=7.5cm]{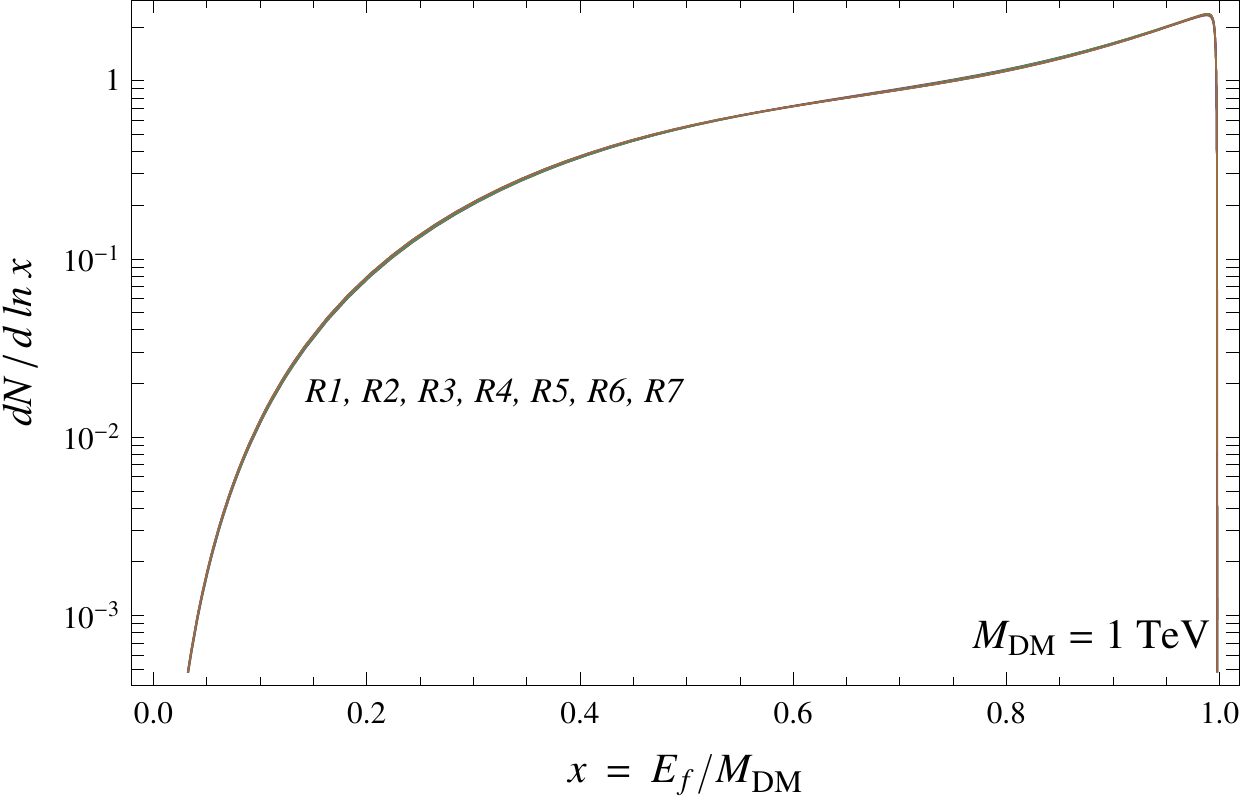}
\hspace{0.5cm}
\includegraphics[width=7.5cm]{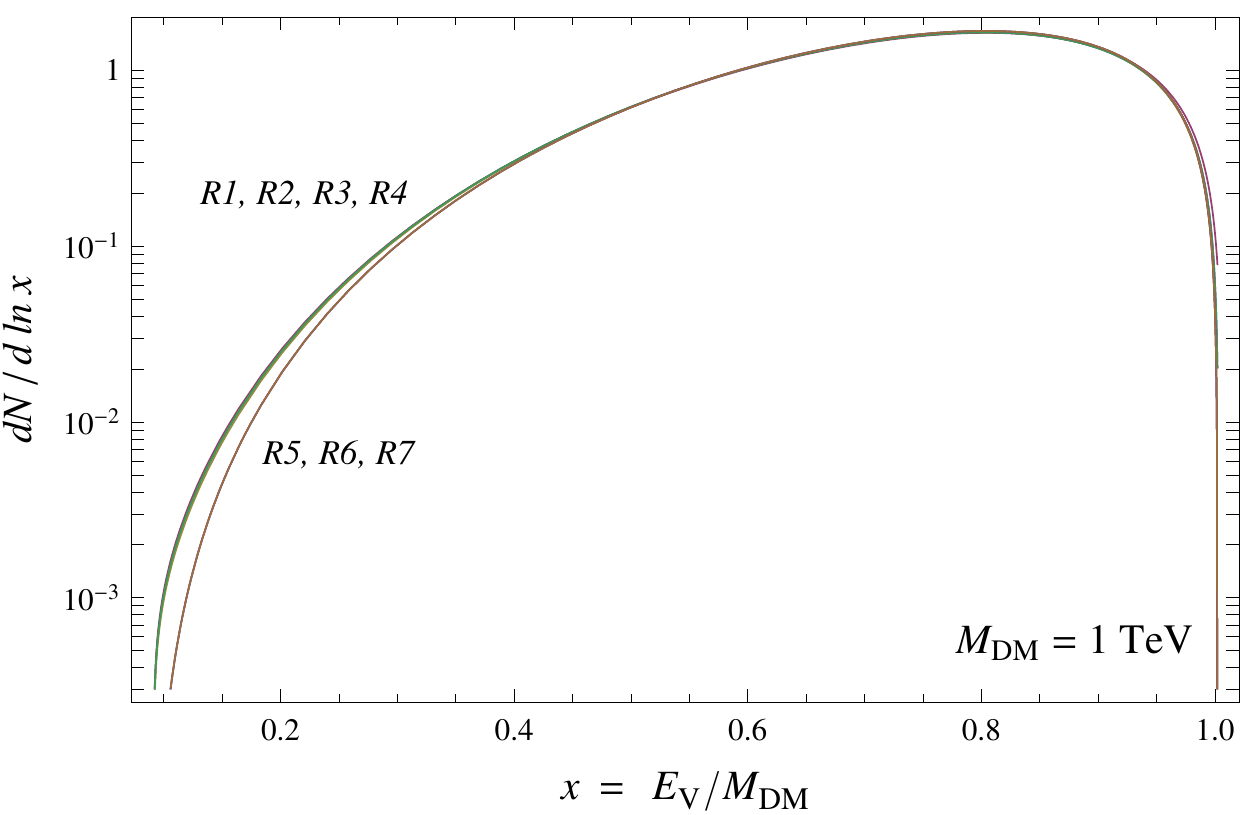}
  \caption{Real scalar DM. \emph{The energy distributions $dN/d\ln(E/M_{\rm DM})$ of the final fermion \emph{(left panel)} and of the final gauge boson \emph{(right panel)},  for the different dimenion-8 operators.
 We  set $c^{(8,s)}_{{\rm R}i}=1$ and $M_{\rm DM}=1$ TeV.
}}
\label{fig:spectraR}
  \end{center}
\end{figure}

For the  case of real scalar DM, we computed the double-differential cross sections
 for the annihilation  process (\ref{eq:MainAnnihilation}) mediated by the dimension-8 operators 
$\mathcal{O}_{{\rm R1}}, \ldots, \mathcal{O}_{{\rm R7}}$ 
listed  in Table \ref{tableOR}
\begin{eqnarray}
\left.v\frac{d^2\sigma}{dydz}\right\vert_{\textrm{R1}} &=&|c_{{\rm R1}}^{(8, s)}|^{2} 
\frac{4 A^2\, M_{\rm DM}^6} {\pi^{3} \Lambda^{8}}
\left[
(1-y-z) (y^{2} + z^{2}) + \frac{m_V^2}{4M_{\rm DM}^2} 
\left[(y+z)^2-2(y+z)+2\right]
\right]\,,\nn\\
&&\\
\left.v\frac{d^2\sigma}{dydz}\right\vert_{\textrm{R2}} &=&|c_{{\rm R2}}^{(8, s)}|^{2} 
\frac{A^2\, M_{\rm DM}^6} {\pi^{3} \Lambda^{8}}
\left[
(1-y-z) (y^{2} + z^{2}) - \frac{m_V^2}{4M_{\rm DM}^2} 
\left[3(y+z)^2-2(y+z+3y z)-2\right]\right.\nn\\
&&\qquad\qquad\qquad\qquad 
\left. - \frac{m_V^4}{8M_{\rm DM}^4} + \frac{m_V^6}{16M_{\rm DM}^6}
\right]\,,\\
\left.v\frac{d^2\sigma}{dydz}\right\vert_{\textrm{R3}} &=&|c_{{\rm R3}}^{(8, s)}|^{2} 
\frac{\,A^2\,M_{\rm DM}^8}{\pi^{3} \Lambda^{8}}
\left[\left(1 - y - z +  \frac{m_V^2}{4 M_{\rm DM}^2} \right) \left( y^{2} + z^{2} - \frac{m_V^{2}}{2 M_{\rm DM}^{2}} \left[ 2(y + z) - 1\right] + \frac{m_V^{4}}{4 M_{\rm DM}^{4}} \right)
\right]\,,\nn\\
&&\\
\left.v\frac{d^2\sigma}{dydz}\right\vert_{\textrm{R4}} &=&
\frac{1}{16}\frac{|c_{{\rm R4}}^{(8, s)}|^{2}}{|c_{{\rm R1}}^{(8, s)}|^{2} }
\left.v\frac{d^2\sigma}{dydz}\right\vert_{\textrm{R1}} 
\,, \\
\left.v\frac{d^2\sigma}{dydz}\right\vert_{\textrm{R5}}  &=& |c_{{\rm R5}}^{(8, s)}|^{2} 
\frac{\, A^2\,M_{\rm DM}^8}{ \pi^{3} \Lambda^{8}}
\left[\left(1 - y - z +  \frac{m_V^2}{4 M_{\rm DM}^2} \right) \left( y^{2} + z^{2} -  \frac{m_V^{2}}{2 M_{\rm DM}^{2}} \right)
\right]\,,\\
\left.v\frac{d^2\sigma}{dydz}\right\vert_{\textrm{R6}}  &=&
\frac{|c_{{\rm R6}}^{(8, s)}|^{2}}{|c_{{\rm R5}}^{(8, s)}|^{2}}
\left.v\frac{d^2\sigma}{dydz}\right\vert_{\textrm{R5}}
\,,\\
\left.v\frac{d^2\sigma}{dydz}\right\vert_{\textrm{R7}}  &=& 
\frac{|c_{{\rm R7}}^{(8, s)}|^{2}}{|c_{{\rm R5}}^{(8, s)}|^{2}}
\left.v\frac{d^2\sigma}{dydz}\right\vert_{\textrm{R5}}
\,.
\end{eqnarray}
They are clearly symmetric under the exchange $y\leftrightarrow z$.
The energy distributions of the fermion and gauge boson, defined as in Eq.~(\ref{dNdx}), 
are shown in Fig.~\ref{fig:spectraR}.
Some of the distributions (e.g. those due to the operators ${\cal O}_{\rm R1}, {\cal O}_{\rm R4}$)
 differ just by an overall factor, which cancels out by normalizing the spectra, and therefore produce the same curves. In the limit $m_V\ll M_{\rm DM}$ the distributions from all different operators become proportional
 to each other.
Therefore, because of the smallness of $m_V/M_{\rm DM}$, 
the differences  in the fermion energy distribution are not visible (left panel of Fig.~\ref{fig:spectraR}).

For convenience, we also report the results for the total cross sections, obtained
after integrating over the kinematical variables

\bea
\left.v\sigma\right\vert_{{\rm R1}}&=&|c_{{\rm R1}}^{(8, s)}|^{2} \frac{ A^{2} M_{\rm DM}^{6}}{60 \pi^{3} \Lambda^{8}} \left[ 8 + 15 \frac{m_{V}^{2}}{M_{\rm DM}^{2}}  + \frac{m_{V}^{4}}{M_{\rm DM}^{4}} \left( 40 - 60 \ln \frac{2 M_{\rm DM}}{m_{V}} \right)
+\mathcal{O}\left(\frac{m_V^6}{M_{\rm DM}^6}\right) \right]\,, \\
\left.v\sigma\right\vert_{{\rm R2}}&=&|c_{{\rm R2}}^{(8, s)}|^{2} \frac{ A^{2} M_{\rm DM}^{6}}{120 \pi^{3} \Lambda^{8}} \left[ 4 + 25 \frac{m_{V}^{2}}{M_{\rm DM}^{2}}  -10 \frac{m_{V}^{4}}{M_{\rm DM}^{4}} \left( 1 + 3 \ln \frac{2 M_{\rm DM}}{m_{V}} \right)
+\mathcal{O}\left(\frac{m_V^6}{M_{\rm DM}^6}\right) \right]\,, \\
\left.v\sigma\right\vert_{{\rm R3}}&=&|c_{{\rm R3}}^{(8, s)}|^{2} \frac{ A^{2} M_{\rm DM}^{6}}{120 \pi^{3} \Lambda^{8}} \left[ 4 - 5 \frac{m_{V}^{2}}{M_{\rm DM}^{2}}  + \frac{m_{V}^{4}}{M_{\rm DM}^{4}} \left( 50 - 30 \ln \frac{2 M_{\rm DM}}{m_{V}} \right)
+\mathcal{O}\left(\frac{m_V^6}{M_{\rm DM}^6}\right) \right]\,, \\
\left.v\sigma\right\vert_{{\rm R4}}&=&{1\over 16}\frac{|c_{{\rm R2}}^{(8, s)}|^{2}}{|c_{{\rm R1}}^{(8, s)}|^{2}}
\left.v\sigma\right\vert_{{\rm R1}}\,,\\
\left.v\sigma\right\vert_{{\rm R5}}&=& |c_{{\rm R5}}^{(8, s)}|^{2} \frac{ A^{2} M_{\rm DM}^{6}}{120 \pi^{3} \Lambda^{8}} \left[ 4  - 15 \frac{m_{V}^{2}}{M_{\rm DM}^{2}} + 5 \frac{m_{V}^{4}}{M_{\rm DM}^{4}} \left( -4 + 6 \ln \frac{2 M_{\rm DM}}{m_{V}}  \right) 
+\mathcal{O}\left(\frac{m_V^6}{M_{\rm DM}^6}\right)\right] \\
\left.v\sigma\right\vert_{{\rm R6}}&=& 
\frac{|c_{{\rm R6}}^{(8, s)}|^{2}}{|c_{{\rm R5}}^{(8, s)}|^{2}}
\left.v\sigma\right\vert_{{\rm R5}}\,,\\
\left.v\sigma\right\vert_{{\rm R7}}&=& 
\frac{|c_{{\rm R7}}^{(8, s)}|^{2}}{|c_{{\rm R5}}^{(8, s)}|^{2}}
\left.v\sigma\right\vert_{{\rm R5}}\,.
\eea
As for the Majorana case, 
we have only shown the leading terms in the expansion in $m_V/M_{\rm DM}$.
The sub-leading terms can  be of the same order as the contributions from higher-dimensional
operators we are neglecting.

\section{Conclusions}
\label{sec:conclusions}

We have considered the annihilation of a self-conjugate DM particle
-- either a Majorana fermion or a real scalar -- into light SM fermions.
In these cases, the simple annihilation DM DM $\to f_L\overline{f}_L$
is helicity suppressed. 
The radiation of electroweak gauge bosons lifts the suppression and opens up an
 $s$-wave channel, independently of the relative velocity of the annihilating DM particles. 
 From the effective operator point of view, 
we found that this effect is encoded by dimension-8 operators.

The dimension-8 operators, despite suffering from a high mass dimensionality, enjoy
no suppression by the relative velocity.
We have found all the dimension-8 operators mediating
 $s$-wave annihilations of DM into SM fermions and a gauge boson, 
 and for each of them we have computed the differential and total cross sections.
 These operators encode the so-called ``virtual internal bremsstrahlung'' effect.

In the Majorana DM case, the two-body annihilation  DM DM $\to f_L\overline{f}_L$ 
 proceeds through a dimension-6 operator and is velocity suppressed ($p$-wave). 
We have shown explicitly and quantitatively 
that the dimension-8 operators actually provide a bigger contribution  to the
annihilation cross section than the dimension-6 operators, for DM masses well above
the weak scale.

For real scalar DM, the $s$-wave  annihilation into massless fermions is forbidden, and becomes
allowed  in the presence of an extra gauge boson in the final state.
We found the dimension-8 operators mediating annihilations into SM fermions and a gauge boson,
which switches on an $s$-wave annihilation process.

In conclusion, our analysis has strengthened the statement that, when dealing with the annihilation of 
 heavy self-conjugate DM today, 
it would be erroneous to neglect the effects of dimension-8 operators.
The results presented in this paper are of interest also when applied to phenomenological analyses, as they provide a model-independent language into which to translate the experimental data, present exclusions and possible future signals.

\section*{Acknowledgements}
We thank Duccio Pappadopulo and Riccardo Rattazzi for useful discussions. 
The work of AT is supported by the
Swiss National Science Foundation under contract 200020-138131
and  by a Marie Curie Early Initial Training Network Fellowship of the European Community's Seventh Framework Programme under contract PITN-GA-2008-237920-UNILHC.
The work of AU is supported by the ERC Advanced Grant n$^{\circ}$ $267985$, ``Electroweak Symmetry Breaking, Flavour and Dark Matter: One Solution for Three Mysteries" (DaMeSyFla).

\appendix

\section{$CP$ transformation properties}
\label{app:CP}

 Under charge conjugation the spinor fields $\psi$ and $\overline{\psi}$ transform as
 \begin{eqnarray}
 \psi&\to&\psi^{\rm C}=\mathcal{C}\overline{\psi}^{T}~,\label{Ctransf1}\\
 \overline{\psi}&\to&\overline{\psi}^{\rm C}=\psi^{T}\mathcal{C}~,\label{Ctransf2}
 \end{eqnarray}
 where the charge conjugation matrix $\mathcal{C}$ obeys $\mathcal{C}^T=\mathcal{C}^{\dag}=\mathcal{C}^{-1}=-\mathcal{C}$.
 Under parity transformation
 \begin{eqnarray}
 \psi&\to& P^{-1}\psi P = i \beta \psi~,\label{Ptransf1}\\
\overline{\psi}&\to& P^{-1}\overline{\psi} P = - i \overline{\psi}\beta\,.\label{Ptransf2}
 \end{eqnarray}
For instance, in the Dirac representation of the $\gamma$-matrices one has
\begin{equation}
 \mathcal{C}= i \gamma^{2} \gamma^{0} = \begin{pmatrix}
   0  &  -i\sigma^2  \\
    -i\sigma^2  & 0 
\end{pmatrix}
\,, \quad
 \beta= \gamma^{0} = 
\begin{pmatrix}
    \textbf{1}   & 0  \\
    0  & -\textbf{1}  
\end{pmatrix}\,.
\end{equation}
The following identities turn out to be useful
 \begin{eqnarray}
 \gamma^{0}  \gamma^{2} \gamma^\mu P_L  \gamma^{2}  \gamma^{0}&=&(\gamma^\mu)^T P_L\,,\\
 \gamma^{0} \gamma^{\mu}P_R\gamma^{0} &=& (-1)^{\mu}\gamma^{\mu} P_L\,,
 \end{eqnarray}
 where
\begin{equation}
(-1)^{\mu}=\left\{\begin{array}{rl}
    +1,& \mu=0 \,,   \\ 
   -1,& \mu=1,2,3\,.
\end{array}\right.
\end{equation}
Using these relations it is easy to prove the following
 transformation properties under $CP$ of operators built out of left-handed fermion fields $f_L$: 
\begin{eqnarray}
CP\left\{\left[\bar f_L\gamma^\mu
\overrightarrow{D}^{\nu} f_L+
\bar f_L\overleftarrow{D}^{\nu}\gamma^\mu f_L\right]\right\}
&=&
(-1)(-1)^\mu(-1)^\nu\,,\\
CP\left\{i\left[\bar f_L\gamma^\mu
\overrightarrow{D}^{\nu} f_L-
\bar f_L\overleftarrow{D}^{\nu}\gamma^\mu f_L\right]\right\}
&=&
(-1)^\mu(-1)^\nu\, \\
CP\left\{\left[\overline{f}_L\overrightarrow{\slashed{D}}\left(\overrightarrow{D}_{\mu}f_L\right)+\left(\overline{f}_L
\overleftarrow{D}_{\mu}\right)\overleftarrow{\slashed{D}}f_L\right]\right\}
&=& (-1)(-1)_{\mu}\,,\\
CP\left\{i\left[\overline{f}_L\overrightarrow{\slashed{D}}\left(\overrightarrow{D}_{\mu}f_L\right)-\left(\overline{f}_L
\overleftarrow{D}_{\mu}\right)\overleftarrow{\slashed{D}}f_L\right]\right\}&=&
(-1)_{\mu}\,,\\
CP\left\{
\left[\overline{f}_L\gamma^{\mu}\overrightarrow{D}_{\rho}\left(\overrightarrow{D}^{\rho}f_L\right)+\left(\overline{f}_L
\overleftarrow{D}_{\rho}\right)\overleftarrow{D}^{\rho}\gamma^{\mu}f_L\right]\right\}
&=& (-1)(-1)^{\mu}\,,\\
CP\left\{
i\left[\overline{f}_L\gamma^{\mu}\overrightarrow{D}_{\rho}\left(\overrightarrow{D}^{\rho}f_L\right)-\left(\overline{f}_L
\overleftarrow{D}_{\rho}\right)\overleftarrow{D}^{\rho}\gamma^{\mu}f_L\right]\right\}
&=& (-1)^{\mu}\,,\\
CP\left\{\overline{f}_L\gamma_{\nu}W^{\mu\nu}f_L\right\}
&=& (-1)_{\mu}\,,\\
CP\left\{\overline{f}_L\gamma_{\nu}\widetilde{W}^{\mu\nu}f_L\right\}
&=& (-1)(-1)_{\mu}\,.
\end{eqnarray}
If $\chi$ is a Majorana fermion field, the bilinears  $\overline{\chi}\gamma^5\chi, 
\overline{\chi}\gamma^{\mu}\gamma^5\chi$
transform under $CP$ as
\begin{eqnarray}
CP\left\{\overline{\chi}\gamma^5\chi\right\}&=&-1 \,,\\
CP\left\{\overline{\chi}\gamma^{\mu}\gamma^5\chi
\right\}&=&(-1)(-1)^{\mu}\,, \label{CPMajoranaside}
\end{eqnarray}
while the bilinears built out of a real scalar $\phi$ transform as
\begin{eqnarray}
CP\left\{\phi\partial_\mu\phi\right\}&=&(-1)^{\mu}\,, \\
CP\left\{\partial_\mu\phi\partial_\nu\phi\right\}
&=&(-1)^\mu(-1)^\nu \,.
\end{eqnarray}

\section{Majorana fermion bilinears}
\label{app:bilinears}

The standard decomposition of a Majorana fermion field in momentum space is
\be
\c(x) = \int \f {d ^ 3 k} {\l ( 2 \pi \r ) ^ 3 \sqrt { 2 E _ k } } \sum _ s \l ( a_s(k) u_s(k) e ^ {- i k\cdot x} 
+ a_s ^ {\dagger}(k) v_s(k) e ^ { i k\cdot x} \r ),
\ee
from which one can compute the matrix element between the initial state with two DM particles 
 and the vacuum, for the bilinears at small velocities. In the  Dirac representation of $\gamma$-matrices we obtain
\be
\l \langle 0 \r | \bar \c \mathcal{O} \c  \l | \c(p^0,\vec p) \c(p^0,-\vec p) \r \rangle
\sim \bar v ( - \vec{p} ) \mathcal {O} u ( \vec {p} ) - \bar v (  \vec{p} ) \mathcal {O} u ( -\vec {p} )\,,
\ee
where the Majorana nature of the initial particles implies to consider the process with spins and momenta
interchanged (and a relative minus sign due to Fermi statistics).
Let us first consider the case where the operator $\mathcal{O}$ is  one of the basis matrices
$\Gamma_A=\{1, \g^5, \g^\mu, \g^5\g^\mu,\sigma^{\mu\nu}\}$. Noting their behaviour under charge conjugation

\be
\mathcal{C}^{-1} \overline{\psi} \Gamma_A \psi \mathcal{C}=\left\{
\begin{matrix}
\overline{\psi} \Gamma_A^T  \psi \quad &(\Gamma_A=1, i \g^5, \g^5\g^\mu)\\
- \overline{\psi} \Gamma_A^T \psi \quad &(\Gamma_A=\g^\mu, \sigma^{\mu\nu})
\end{matrix}
\right. \, ,
\label{chargeconj}
\ee
we find that
\be
 \bar v (  \vec{p} ) \Gamma_A u ( -\vec {p} ) =
 \left\{
\begin{matrix}
-\bar v ( - \vec{p} ) \Gamma_A u ( \vec {p} )  \quad &(\Gamma_A=1, i \g^5, \g^5\g^\mu)\\
+\bar v ( - \vec{p} ) \Gamma_A u ( \vec {p} )  \quad &(\Gamma_A=\g^\mu, \sigma^{\mu\nu})
\end{matrix}
\right.\,,
\ee
hence the vector and tensor operators do not contribute to the annihilation of Majorana particles --
as it is well known -- and it is sufficient to compute the bilinears
for scalar, pseudo-scalar and pseudo-vector interactions
\be
\begin{array}{lllll}
\bar v(-\vec p)u(\vec p) & = & -2 \eta^\dag (\vec\sigma\cdot\vec p) \xi & \sim & v^1\,,\nn\\
\bar v(-\vec p)\gamma^5 u(\vec p) & = & - (E + M) \, \eta^\dag \xi  & \sim & v^0\,,\nn\\
\bar v(-\vec p)\gamma^0\gamma^5 u(\vec p) & = & (E + M) \, \eta^\dag \xi & \sim & v^0\,, \nn\\
\bar v(-\vec p)\gamma^i\gamma^5 u(\vec p) & = & - 2i  \eta^\dag (\vec p\times \vec\sigma)^i \xi & \sim & v^1\,.
\end{array}
\ee
The bilinears $\bar u(\vec p) \Gamma_A v(-\vec p)$ can be obtained from the previous ones by
\be
\bar u(\vec p) \Gamma_A v(-\vec p)=\l[\bar v(-\vec p) (\gamma^0 \Gamma_A^\dag \gamma^0) u(\vec p)\r]^{\dagger}\,,
\ee
such that, for example
\bea
\bar u(\vec p)\gamma^0\gamma^5 v(-\vec p) &=& (E + M) \, \xi^\dag \eta \,, \\
\bar u(\vec p)\gamma^i\gamma^5 v(-\vec p) &=& 2i (\vec p\times \vec\sigma)^i \, \xi^\dag\eta\,.
\eea

\noindent
If the operator $\mathcal{O}$ contains one derivative, we have
\bea
\bar \c \g ^ 5 \partial_{0} \c & \sim & v^0, \nn\\
\bar \c \g ^ 5 \partial_{i} \c & \sim & v, \nn\\
\bar \c \g ^ 5 \g^{0} \partial_{0} \c & \sim & v^0 , \nn \\
\bar \c \g ^ 5 \g ^ 0 \partial_{i} \c & \sim & v , \nn \\
\bar \c \g ^ 5 \g^{i} \partial_{j} \c & = & 0.
\eea
The generalization to an arbitrary number of derivatives is straightforward.

\section{Useful Identities}
 \label{app:identities}

By a direct computation we find the following identity involving one covariant derivative
\begin{equation}\label{eq:IDfirstderivative}
\partial_{\nu}\left(\overline{f}_L\gamma^{\mu}f_L\right)=
\overline{f}_L\gamma^{\mu}(\overrightarrow{D}_{\nu}f_L)+(
\overline{f}_L\overleftarrow{D}_{\nu})\gamma^{\mu}f_L ~.
\end{equation}
A particular case is the contraction $\mu=\nu$, which gives, using the equations of motion 
\begin{equation}\label{eq:ID1bis}
\partial_{\mu}\left(\overline{f}_L\gamma^{\mu}f_L\right)=0~.
\end{equation}
For two covariant derivatives we find
\begin{equation}\label{eq:2derivative}
\partial_{\rho}\partial_{\nu}\left(\overline{f}_L\gamma^{\mu}f_L\right)=
\overline{f}_L\gamma^{\mu}\overrightarrow{D}_{\rho}(\overrightarrow{D}_{\nu}f_L)+
(\overline{f}_L\overleftarrow{D}_{\rho})\gamma^{\mu}(\overrightarrow{D}_{\nu}f_L)+
(\overline{f}_L\overleftarrow{D}_{\nu})\gamma^{\mu}(\overrightarrow{D}_{\rho}f_L)+
(\overline{f}_L\overleftarrow{D}_{\nu})\overleftarrow{D}_{\rho}\gamma^{\mu}f_L~.
\end{equation}
A useful relation can be extracted by contracting  $\rho=\mu$
\begin{equation}
\partial_{\mu}\partial_{\nu}\left(\overline{f}_L\gamma^{\mu}f_L\right)=
\overline{f}_L\overrightarrow{\slashed{D}}(\overrightarrow{D}_{\nu}f_L)+
(\overline{f}_L\overleftarrow{D}_{\nu})\overleftarrow{\slashed{D}}f_L ~.
\end{equation}
However, exchanging the partial derivatives on the left hand side, and using Eq. (\ref{eq:ID1bis}) we find
\begin{equation}\label{eq:2d}
\overline{f}_L\overrightarrow{\slashed{D}}(\overrightarrow{D}_{\nu}f_L)+
(\overline{f}_L\overleftarrow{D}_{\nu})\overleftarrow{\slashed{D}}f_L=0~.
\end{equation}
 Eq.~(\ref{eq:2derivative}) shows that the action of a partial derivative is equivalent to the action of the corresponding covariant derivative on each single term. This relation is crucial to relate  operators containing quadratic 
 ($\supset \phi^2$) and derivative ($\supset \phi\partial_\mu\phi$) terms of the scalar field. 
 Notice that this kind of relation remains true also for differences of SM operators instead of sums.
 By direct computation we find
 \begin{eqnarray}\label{eq:relation3}
&&\partial_{\rho}\left[
\overline{f}_L\gamma^{\mu}(\overrightarrow{D}_{\nu}f_L)-(
\overline{f}_L\overleftarrow{D}_{\nu})\gamma^{\mu}f_L\right]=\nonumber\\
&&\overline{f}_L\gamma^{\mu}\overrightarrow{D}_{\rho}(\overrightarrow{D}_{\nu}f_L)+
(\overline{f}_L\overleftarrow{D}_{\rho})\gamma^{\mu}(\overrightarrow{D}_{\nu}f_L)-
(\overline{f}_L\overleftarrow{D}_{\nu})\gamma^{\mu}(\overrightarrow{D}_{\rho}f_L)-
(\overline{f}_L\overleftarrow{D}_{\nu})\overleftarrow{D}_{\rho}\gamma^{\mu}f_L~.
 \end{eqnarray} 
Similar relations hold for three covariant derivatives
\begin{eqnarray}\label{eq:3derivatives}
\partial_{\sigma}\partial_{\rho}\partial_{\nu}\left(\overline{f}_L\gamma^{\mu}f_L\right) & = &
\overline{f}_L\gamma^{\mu}\overrightarrow{D}_{\sigma}\overrightarrow{D}_{\rho}(\overrightarrow{D}_{\nu}f_L)+
(\overline{f}_L\overleftarrow{D}_{\sigma})\gamma^{\mu}\overrightarrow{D}_{\rho}(\overrightarrow{D}_{\nu}f_L)+
(\overline{f}_L\overleftarrow{D}_{\rho})\gamma^{\mu}\overrightarrow{D}_{\sigma}(\overrightarrow{D}_{\nu}f_L) \nonumber\\
& + &
(\overline{f}_L\overleftarrow{D}_{\rho})\overleftarrow{D}_{\sigma}\gamma^{\mu}(\overrightarrow{D}_{\nu}f_L)+
(\overline{f}_L\overleftarrow{D}_{\nu})\gamma^{\mu}\overrightarrow{D}_{\sigma}(\overrightarrow{D}_{\rho}f_L)+
(\overline{f}_L\overleftarrow{D}_{\nu})\overleftarrow{D}_{\sigma}\gamma^{\mu}(\overrightarrow{D}_{\rho}f_L) \nonumber\\
&+&
(\overline{f}_L\overleftarrow{D}_{\nu})\overleftarrow{D}_{\rho}\gamma^{\mu}(\overrightarrow{D}_{\sigma}f_L)+
(\overline{f}_L\overleftarrow{D}_{\nu})\overleftarrow{D}_{\rho}\overleftarrow{D}_{\sigma}\gamma^{\mu}f_L ~.
\end{eqnarray}
This cumbersome identity simplifies contracting $\rho=\mu$; using the same argument previously exploited we find
\begin{eqnarray}
\partial_{\sigma}\partial_{\mu}\partial_{\nu}\left(\overline{f}_L\gamma^{\mu}f_L\right)&=&
\overline{f}_L \overrightarrow{D}_{\sigma}\overrightarrow{\slashed{D}}(\overrightarrow{D}_{\nu}f_L)+(\overline{f}_L\overleftarrow{D}_{\sigma})\overrightarrow{\slashed{D}}(\overrightarrow{D}_{\nu}f_L)\nonumber\\
&+&(\overline{f}_L\overleftarrow{D}_{\nu})\overleftarrow{\slashed{D}}\gamma^{\mu}(\overrightarrow{D}_{\sigma}f_L)+
(\overline{f}_L\overleftarrow{D}_{\nu})\overleftarrow{\slashed{D}}\overleftarrow{D}_{\sigma}\gamma^{\mu}f_L ~,
\end{eqnarray}
which is useful to find dependencies among seemingly independent operators.



\end{document}